# IT based social media impacts on Indonesian general legislative elections 2014

**Leon Andretti Abdillah[a]**

[a] Bina Darma University, Computer Science Faculty, Information Systems Study Prorgam, Palembang 30264, INDONESIA
leon.abdillah@yahoo.com

**Abstract.** The information technology's applications in cyberspace (the internet) are currently dominated by social media. The author investigates and explores the advantages of social media implementation of any political party in Indonesian general legislative elections 2014. There are twelve national political parties participating in the election as contestants plus three local political parties in Aceh. In this researh, auhtor focus on national politial parties only. The author visited, analyzed, and learnt the social media used by the contestants. Those social media are : 1) Facebook, 2) Twitter, and 3) YouTube. Author also compares the popularity of political parties on social media with the results of a real count. Then Author can discuss : 1) the impact of social media on political parties, 2) social media as a brand of political parties, 3) social media as political presentation, and 4) social media as virtual society. The results of this study indicate that Facebook is still a social media application that received high attention by the voters on a campaign of political parties. Indonesian's legislative elections won by parties that are using social media as part of their campaigns.

**Keywords:** Social media, political campaigns, Indonesian legislative elections.

## 1 INTRODUCTION

Social media become one of the top internet application recently. These applications grow exponentially and attract millions concerns from virtual users. Social media are being discussed and analyze not only by industry and academic, but also involved politicians. Recently, social media have been used for personal online interaction, academic and education (Abdillah, 2013; Selwyn, 2009), product promotion (Rahadi & Abdillah, 2013), knowledge and information sharing (Abdillah, 2014), politics (Abdillah, 2014; Baumgartner & Morris, 2010), or activities (Boyd & Ellison, 2007). The rapid development of online social networks has tremendously changed the way of people to communicate with each other (Bi, Qin, & Huang, 2008). This article covers the topic of social media impacts on involving citizens in supporting e-democratic activity like general legislative elections or presidential campaigns.

Indonesia become the third largest democratic country after USA, and India. In Indonesia there are two steps of general national elections: 1) legislative elections, and 2) presidential elections. Legislative elections conducted to elect representatives in Senayan, House of representatives. The representatives come from a variety of political parties. In the 2014 election, there are twelve national political parties and three local political parties from Aceh. Research discussion in this article focus on : 1) legislative elections, 2) twelve national political parties, and 3) social media used by the national political parties.

In legislative elections, every political party must prepare and promote candidates for legislative council. Normally, in the campaign or promotion periods every political party will disseminate their politicians using various of conventional media, such as: 1) televisions, 2) news papers, 3) radios, 4) banners, etc.

Along with advances in information technology, political parties in Indonesia have also adopted social media in political campaigns and activities. This condition occured because of citizens already friendly with the social media and become bored with conventional campaigns. Adding new media to old electoral politics will entice new and younger voters to greater participation (Xenos & Foot, 2008), because there are relationships between Facebook use and students' life satisfaction, social trust, civic engagement, and political participation (Valenzuela, Park, & Kee, 2009). On the other hand, information networks not easily controlled by the state and coordination tools that are already embedded in trusted networks of family and friends (Howard & Hussain, 2011). Organizations such as political parties are trying to keep up with this changing environment (Effing, van Hillegersberg, & Huibers, 2011). Another reason is based on one common characteristic among social media sites is that they tend to be free and are therefore widely accessible across socioeconomic classes (Joseph, 2012). Educated and well inform people less trust to billboards or banners, but they have more confidence or rather believe in the words of friends or colleagues in social media (Sugiarto, 2014).

The greatest phenomenon is Barack Obama's campaign in 2008. The successful use of social media in the US presidential campaign of Barack Obama (Tumasjan, Sprenger, Sandner, & Welpe, 2010) has established Twitter, Facebook, MySpace, and other social media as integral parts of the political campaign toolbox and how they have affected users' political attitudes and behaviors (Zhang, Johnson, Seltzer, & Bichard, 2010). Another success story is from Indonesia, Jokowi & Ahok, new Governor and deputy as winners of Jakarta Governor Election in the 2012s suggest political marketing strategy is an effective key to success (Ediraras, Rahayu, Natalina, & Widya, 2013).

Table 1. Facebook user distribution based on SocialBakers

| No | Country | Facebook user based on age | | Facebook user based on gender | |
|---|---|---|---|---|---|
| | | The Largest | The Second Largest | Male | Female |
| 1 | USA | 25-34 | 18-24 | 46% | 54% |
| 2 | India | 18-24 | 25-34 | 76% | 24% |
| 3 | Brazil | 18-24 | 25-34 | 47% | 53% |
| 4 | **Indonesia** | **18-24** | **25-34** | **59%** | **41%** |
| 5 | Mexico | 18-24 | 25-34 | 50% | 50% |

As of April 2014, Indonesia is the world's fourth position in terms of the number of facebook users(SocialBakers, 2014) after USA, India, and Brazil.

Young adults (18-24) people dominate Facebook users in Indonesia followed by the users in the age of 25-34 (table 1).There are 59% male users and 41% female users in Indonesia, compared to 46% and 54% in USA, 76% and 24% in India, 47% and 53% in Brazil, and 50% and 50% in Mexico.

In this article, author would like to discuss about successful story in social media based political campaign in Indonesia. Author extends this article by adding new data from real count result, add some sections, and add one social media, YouTube. The rest of this paper will discuss research methods, results and discussions, and conclude with conclusions and the direction for next study.

Information about political party contestants, national poltical party's website such as: Facebook, Twitter, and, YouTube, etc will be discussed in next section.

## 2 RESEARCH METHODS

This article is a continuation of the previous article analyzes (Abdillah, 2014) conducted by the author. In this article author add one social media, YouTube, for the observation. Then in this article author observes the features of political parties' social media, such as: 1)

Facebook, 2) Twitter, and 3) YouTube. Author explores political parties' social media to check their activities. Author also gathers valuable information from KPU for the real count results (2009 and 2014).

### 2.1 Indonesian Political Parties for General Legislative Elelctions 2014

On elections in this era of reform involves twelve political parties consisting of eleven existing political parties and one new, Pnasdem, plus three local political parties in Aceh. In this article, author will not discuss three others local political parties in Aceh.

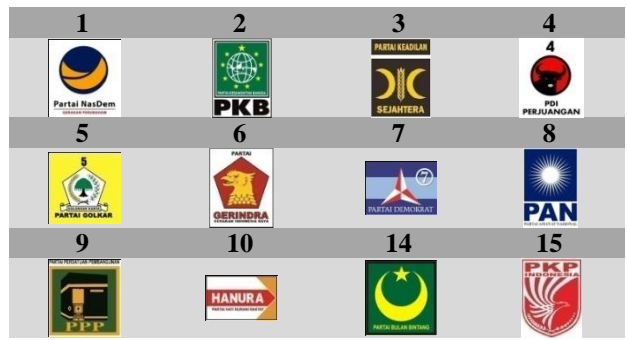

Fig. 1. Political Parties' Facebook Like's Statistics

### 2.2 Indonesian Political Parties's Social Media Accounts

This study only investigate three of the most popular social media at the moment. Table 2 shows the list of all social media accounts for political parties: 1) Facebook Page, 2) Twitter account, and 3) YouTube address.

Table 2. Polical Parties Social Media Accounts

| No | Political Party | Facebook Page | Twitter | YouTube |
|---|---|---|---|---|
| 1 | PNasdem | - | @NasDem | - |
| 2 | PKB | pkb2pkb | @PKB_News_Online | - |
| 3 | PKS | HumasPartaiKeadilanSejahtera | @PKSejahtera | http://www.youtube.com/user/konpress |
| 4 | PDIP | DPP.PDI.Perjuangan | @PDI_Perjuangan | - |
| 5 | PGolkar | DPPPGolkar | @Golkar2014 | - |
| 6 | PGerindra | gerindra | @Gerindra | http://www.youtube.com/user/GerindraTV |
| 7 | PDemokrat | pdemokrat | @PDemokrat | http://www.youtube.com/user/demokrattv |
| 8 | PAN | amanatnasional | @official_PAN | http://www.youtube.com/user/OfficialDPPPAN |
|   |   |   |   | http://www.youtube.com/user/pan230898 |
| 9 | PPP | pppdpp | @DPP_PPP | - |
| 10 | PHanura | hanura.official | @hanura_official | - |
| 11 | PBB | DPP-Partai-Bulan-Bintang-wwwbulan-bintangorg/114716555303039 | @DPPBulanBintang | - |
| 12 | PKPI | PKPI.MediaCenter | @sobatbangyos | - |

### 2.3 Statistics from Online Media

To enrich the analysis from every social media that involved in political parties, author gathers information from some of statistical online media, such as SocialBakers. SocialBakers is an important online statistics tool for researchers who want to find out the social media condition from every country. SocialBakers works with four major social at the moment: 1) Facebook, 2) Twitter, 3) YouTube, and 4) Linkedin.

SocialBakers is a user friendly social media analytics platform which provides a leading global solution that allows brands to measure, compare, and contrast the success of their social media campaigns with competitive intelligence (SocialBakers, 2014). SocialBakers also provide eleven clasifications, such as: 1) By Country, 2) Pages, 3) Brands, 4) Media, 5) Entertainment, 6) Sport, 7) Celebrities, 8) Society, 9) Community, 10) Places, and 11) Apps & Developers.

## 3 RESULTS AND DISCUSSIONS

In this section, author would like to show the results from real count by KPU, then followed by information about political parties' social media. Author also discusses the impact of social media.

### 3.1 Real Count Results

In the last article, author uses quick count results from some survey's institutions. In this article author uses real count data from legal institutions, KPU. Table 3 shows the real count results from the elections 2009 and 2014.

Based on 2014 real counts result, there are new winner in this election : 1) **PDIP**, 2) **PGolkar**, 3) **PGerindra**, 4) PDemokrat, 5) PKB, 6) PAN, 7) PNasdem, 8) PPP, 9) PKS, 10) PHanura, 11) PBB, 12) PKPI.

Table 3. Real Count Results in Indonesia General Election 2014

| Rank | Political Party | Elections 2009 | Elections 2014 | Change |
|---|---|---|---|---|
| 1 | PDIP | 14.03% | 18.95 % | + (↑) |
| 2 | PGolkar | 14.45% | 14.75 % | + (↑) |
| 3 | PGerindra | 4.46% | 11.81 % | + (↑) |
| 4 | PDemokrat | 20.85% | 10.19 % | - (↓) |
| 5 | PKB | 4.94% | 9.04 % | + (↑) |
| 6 | PAN | 6.01% | 7.59 % | + (↑) |
| 7 | PKS | 7.88% | 6.79 % | - (↓) |
| 8 | PNasdem | - | 6.72 % | + (↑) |
| 9 | PPP | 5.32% | 6.53 % | + (↑) |
| 10 | PHanura | 3.77 % | 5.26 % | + (↑) |
| 11 | PBB | 1.79 % | 1.46 % | - (↓) |
| 12 | PKPI | 0.90 % | 0.91 % | + (↑) |

Among those twelve political parties, only one new political party joins the election of 2014, PNasdem. Eight political parties get better voices in the election 2014, and three others political parties get lower voices then the elections of 2009.

The increasing number of voice revenue for 2014 elections won by the party PGerindra (+7.35%), followed by PDIP (+4.92%) and PKB (+4.1%). Author doesn't include PNasdem because this political party just join the election in 2014. PGolkar successfully defended his party's ranking in 2014 elections as runner up in the election 2009 & 2014. Figure 2 shows the tabular bar for the real count results of the elections in 2009 and 2014.

Unfortunately, in this election, none of the parties get more votes than 20%, as a condition to nominate candidates for president himself. So that each party must set stategy for Formatting a coalition with a number of parties. Author classifies the group of political parties in the election 2014: 1) three big political parties (PDIP, PGolkar, PGerindra), 2) three medium political parties (PDemokrat, PKB, PAN), 3) four small political parties (PNasdem, PKS, PPP, PHanura), and 4) two political parties which threatened not to go to parliament.

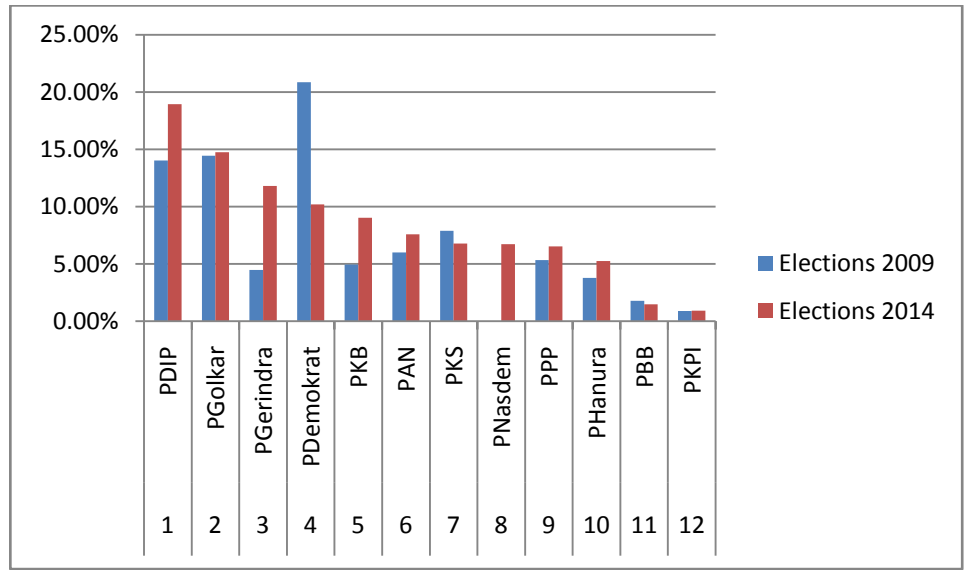

Fig. 2. Political Parties' Real Count Results (2009 & 2014)

## 3.2 Political Parties' Facebook Page

The first social media analyzed in this article is Facebook. The Facebook connectivity help the group to build a political party to further back up their main figure in the forecast Presidential candidacy (Murti, 2013). Table 4 shows the popularity of every political partiy in general election April 2014.

Table 4. Political Parties's Facebook Like

| No | Political Party | Like (April, before real Count) | Like (May, after real count) | Talking about this | Most Popular Age Group |
|---|---|---|---|---|---|
| 1 | P.Nasdem | - | - | - | - |
| 2 | PKB | 6.164 K | 6.333 K | 0.319 K | 25-34 |
| 3 | PKS | 40.073 K | 40.552 K | 1.729 K | 25-34 |
| 4 | PDIP | 319.000 K | 453.559 K | 273.756 K | 18-24 |
| 5 | P.Golkar | 4.355 K | 4.394 K | 0.007 K | 18-24 |
| 6 | P.Gerindra | 2500.000 K | 2650.000 K | 326.304 K | 18-24 |
| 7 | P.Demokrat | 25.075 K | 28.000 K | 1400.000 K | 18-24 |
| 8 | PAN | 38.228 K | 45.000 K | 7.3K | 18-24 |
| 9 | PPP | 3.388 K | 3.596 K | 0.225 K | 25-34 |
| 10 | P.Hanura | 562.000 K | 563.918 K | 0.629 K | 18-24 |
| 11 | PBB | 2.198 K | 2.211 K | 0.009K | 25-34 |
| 12 | PKPI | 4.410 K | 4.095 K | 0.007K | 25-44 |

Author found that the popularity of political party in the general election 2014 based on the Facebook page political parties, there are three parties managed to get "Like" the most. Those political parties are : 1) PGerindra, 2) PHanura, and 3) PDIP. Among the three most popular parties, there are PGerindra and PDIP included in the election winners. This research also confirm that dominanat Facebook's users that like political parties are young adults (19-24) followed by adults (25-34).

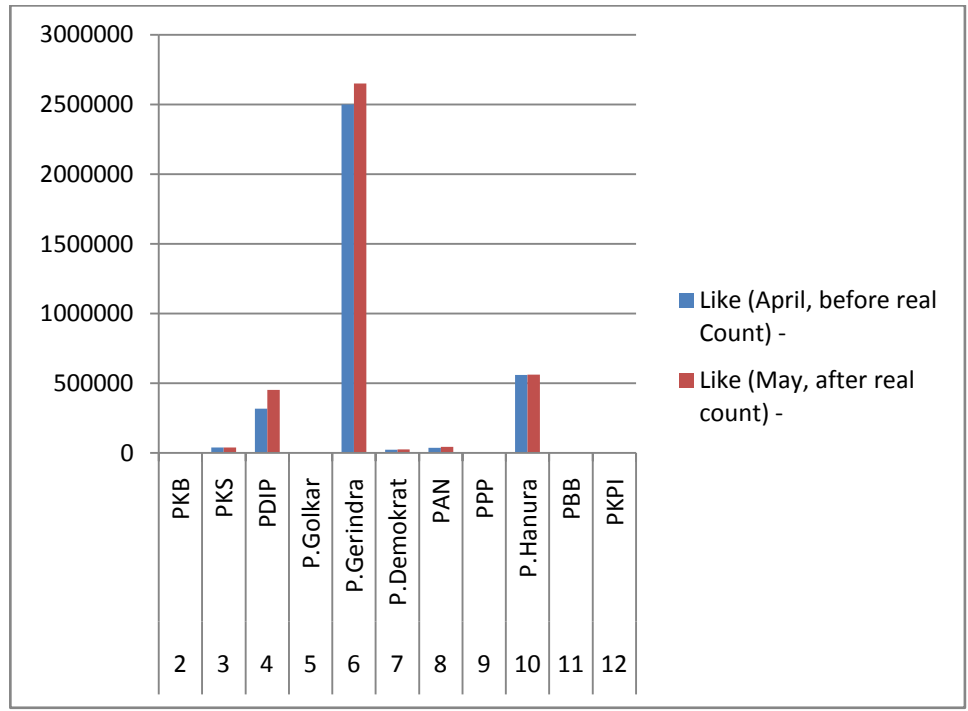

Fig. 3. Political Parties' Facebook Like's Statistics

## 3.3 Political Parties' Twitter

Twitter is a microblogging website where users read and write millions of short messages on a variety of topics every day (Tumasjan, et al., 2010). The analysis suggests that politicians are attempting to use Twitter for political engagement, though some are more successful in this than others (Grant, Moon, & Busby Grant, 2010). All of national political parties have official website (Table 1) and twitter account (Table 2).

Table 5. Political Parties's Twitter's Tweets, Following, and Followers

| No | Political Party | Tweets (April) | Tweets (May) | Following (April) | Following (May) | Followers (April) | Followers (May) |
|---|---|---|---|---|---|---|---|
| 1 | P.Nasdem | 17.600 K | 19.865 K | 0.669 K | 0.688 K | 20.900 K | 22.582 K |
| 2 | PKB | 2.898 K | 2.991 K | 1.704 K | 1.704 K | 4.050 K | 4.270 K |
| 3 | PKS | 18.900 K | 19.038 K | 0.275 K | 0.275 K | 105.000 K | 109.272 K |
| 4 | PDIP | 21.300 K | 23.606 K | 0.658 K | 0.704 K | 58.400 K | 64.746 K |
| 5 | P.Golkar | 9.680 K | 0.074 K | 0.493 K | 0.027 K | 2.329 K | 0.017 K |
| 6 | P.Gerindra | 47.600 K | 48.930 K | 2.160 K | 3.173 K | 143.000 K | 149.711 K |
| 7 | P.Demokrat | 4.008 K | 4.148 K | 0.870 K | 0.872 K | 18.200 K | 19.222 K |
| 8 | PAN | 6.022 K | 6.243 K | 0.605 K | 0.605 K | 4.745 K | 5.216 K |
| 9 | PPP | 4.424 K | 4.511 K | 0.076 K | 0.073 K | 2.953 K | 3.155 K |
| 10 | P.Hanura | 1.256 K | 1.257 K | 0.030 K | 0.030 K | 1.866 K | 1.937 K |
| 11 | PBB | 0.172 K | 0.172 K | 0.014 K | 0.014 K | 0.774 K | 0.806 K |
| 12 | PKPI | 2.301 K | 2.301 K | 1.376 K | 1.377 K | 1.364 K | 1.385 K |

Almost twitter accounts from all political parties included the abreviation of their political party's name except PKPI which use @sobatbangyos. Table 4 shows political partiy's twitter account. Table 5 shows the numbers of tweets, following, and followers from every political parties. All of political parties have twitter account. In this manuscript, author will analyze twitter features of every political party based on : 1) Twitters' tweets, 2) following, and 3) followers.

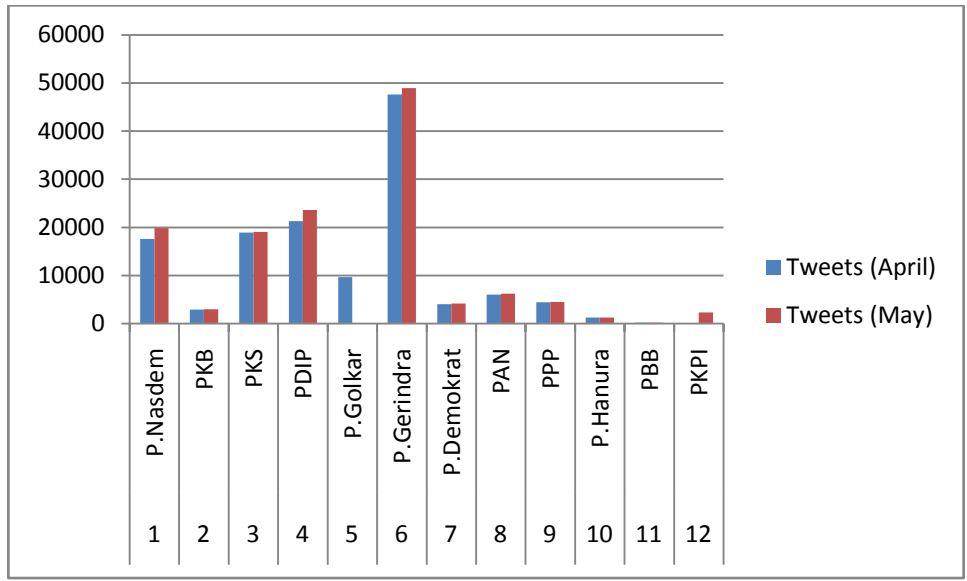

Fig. 4.a. Political Parties' Twitters Tweets

Twitter, every user can publish short messages with up to 140 characters, so-called "tweets" (Tumasjan, et al., 2010), which are visible on a public message board of the website or through third-party applications. Figure 4.a shows the numbers of tweets of every political party. PGerindra was the most informative political party with 48,9 K tweets followed by PDIP (23.6 K) and PNasdem (19.8 K).

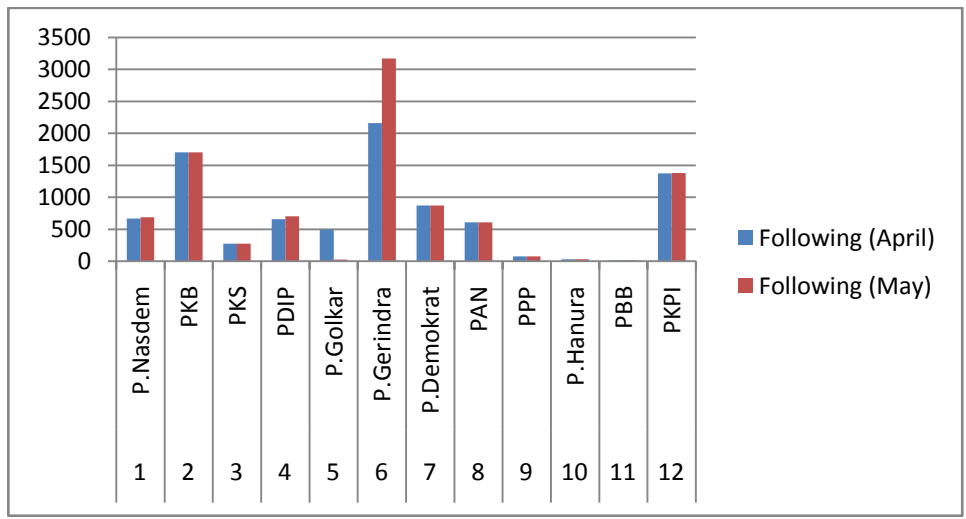

Fig. 4.b. Political Parties' Twitters Following

Figure 4.b shows the numbers of following of every political party. P.Gerindra was the most kindest political party that has been following 2,16 K others twitter accounts followed by PKB and PKPI.

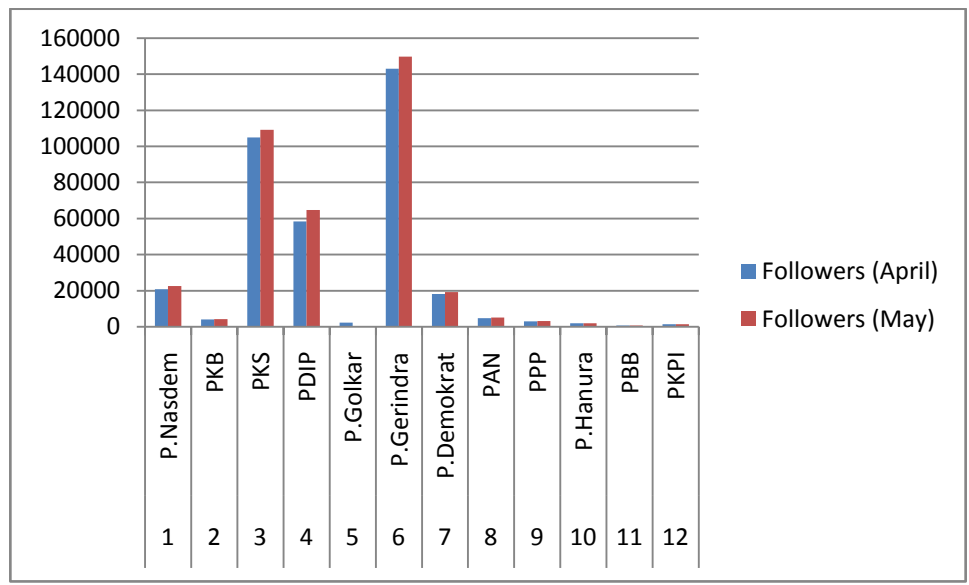

Fig. 4.c. Political Parties' Twitters Followers

Table 4.c shows the number of twitter followers of political parties. Based on these results, PGerindra occupies the first position, followed by PKS and PDIP. These results also show that PGerindra followers increase significantly after the announcement of real count result.

### 3.4 Political Parties' YouTube

Beside two most popular socia media above, national political patries in Indonesia also use video-sharing site YouTube for the third alternative for their online campaigns. Table 6 shows national party's YouTube account with total subscribers. YouTube as one of video-sharing websites or free distribution channel provides a low-cost alternative for many political candidates to use in their campaigns (English, Sweetser, & Ancu, 2011).

Table 6. Political Parties's YouTube's Subscribers

| No | Political Party | YouTube | Subscribers (April) | Subscribers (May) |
|---|---|---|---|---|
| 1 | P.Nasdem | - | - | - |
| 2 | PKB | - | - | - |
| 3 | PKS | http://www.youtube.com/user/konpress | 5.768 K | 6.078 K |
| 4 | PDIP | - | - | - |
| 5 | P.Golkar | - | - | - |
| 6 | P.Gerindra | http://www.youtube.com/user/GerindraTV | 5.023 K | 6.108 K |
| 7 | P.Demokrat | http://www.youtube.com/user/demokrattv | 0.130 K | 0.152 K |
| 8 | PAN | http://www.youtube.com/user/OfficialDPPPAN | 0.011 K | 0.012 K |
|   |   | http://www.youtube.com/user/pan230898 | 0.030 K | 0.030 K |
| 9 | PPP | - | | |
| 10 | P.Hanura | - | | |
| 11 | PBB | - | | |
| 12 | PKPI | - | | |

YouTube as one of video-sharing websites or free distribution channel provides a low-cost alternative for many political candidates to use in their campaigns (English, Sweetser, & Ancu, 2011). In Obama's campaign, using YouTube more for media and message than for mobilization (Nielsen, 2011), but YouTube videos were successful in mobilizing younger voters like never before (McKinney & Rill, 2009).

The current trend in the use of YouTube as a medium for political campaigns has penetrated into Indonesia although not as optimal as practiced by Obama. Currently, some political parties have been used YouTube as political virtual channel and succeed to atract thousands attention from online users. At least there are four political parties used YouTube, namely: 1) PGerindra, 2) PKS, 3) PDemokrat, and 4) PAN.

Even PKS succeed to get the highest subscribers in YouTube, unfortunately PKS failed to get many votes in the general legislative elections 2014. Among the four political parties, PGerindra not only get a good number of subscribers, but PGerindra also won the third place based on real count results from KPU.

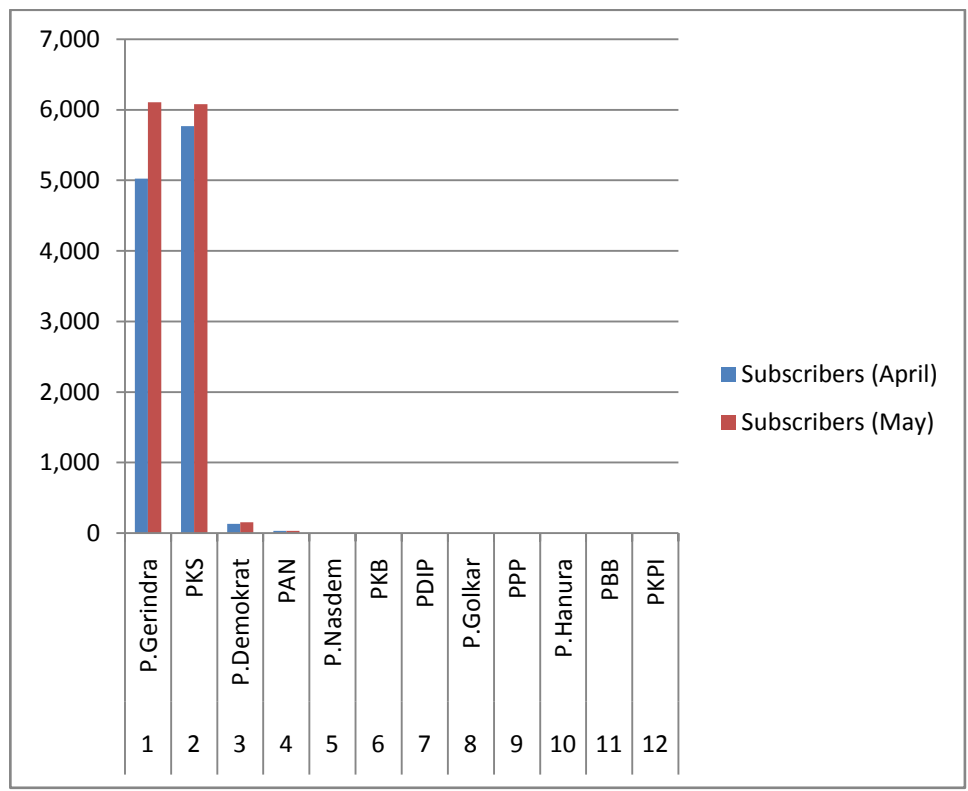

Fig. 5. Political Parties' YouTube Subscribers

## 3.5 Social Media as Political Party Brand

The term of brand refers to image and reputation is more or less right (Scammell, 2007). Social media can be used to promote legislative candidates to all prospective voters easily. Every political party has provided a number of social media for their loyalists as well as to capture new enthusiasts of the virtual user. In general legislative elections of 2014, Indonesian poitical parties are still dependent on the reputation of its chairperson or leader icon. For example, on Jokowi's political branding in governor election campaign is twitter social media (Wulan, Suryadi, & Dwi Prasetyo, 2014).

SocialBakers page report that political leaders and political parties get a very high attention in the social media of Facebook as community. Based on Figure 6, Prabowo and his political party, PGerindra, managed to occupy the top three social media community page in Facebook. Other political figures who entered the top ten are: SBY (no.2), BJ Habibie (no.7), and Muh. Jusuf Kalla (no.8). While the political parties who entered the top ten only PGerindra (no.3).

### 3.6 Social Media as Political Presentation

Social media platforms give politicians access to millions of users and offer the capacity to build a sense of camaraderie and connection with a wide constituency (Crawford, 2009). As political communication sources, social media are a recent phenomenon (Kushin & Yamamoto, 2010). Author browsed every political parties in April 2014. Author found that every political parties has been accupied the political parties' Facebook page with some features.

Table 7. Political Parties's Facebook Page Views

| No | PP | Logo | Order No | Head | Quote, Slogan, TagLine | Others | Website |
|----|----|------|----------|------|------------------------|--------|---------|
| 1 | P.Nasdem | - | - | | | | - |
| 2 | PKB | ✓ | ✓ | - | - | | ✓ |
| 3 | PKS | ✓ | - | ✓ | ✓ | Coalition | - |
| 4 | PDIP | ✓ | - | ✓ | - | Soekarno | - |
| 5 | P.Golkar | ✓ | - | - | ✓ | - | - |
| 6 | P.Gerindra | ✓ | - | ✓ | ✓ | - | - |
| 7 | P.Demokrat | ✓ | ✓ | ✓ | ✓ | - | - |
| 8 | PAN | ✓ | ✓ | ✓ | ✓ | - | - |
| 9 | PPP | ✓ | ✓ | - | ✓ | - | - |
| 10 | P.Hanura | ✓ | ✓ | - | ✓ | KPU | - |
| 11 | PBB | ✓ | - | ✓ | - | Simpatisan | ✓ |
| 12 | PKPI | ✓ | ✓ | ✓ | ✓ | Pemilu.com | - |
| | Persentage | 100% | 50% | 58.3% | 66.67% | 41.67% | |

Mayl 2014

Author visited every political parties' facebook account. Author found that every political party has different way to present their existness via facebook. Table 8 show the summary of political parties appereance in facebook.

Table 7 shows the summary of political parties' facebook pages' features. In 2014 general elections, all of political parties (100%) dispaly their logo, half of them (50%) showing the serial number of his party, 58% display photos political party chairman, 66,67% have quote/slogan/tagline, and 41.67% showing other things.

Unfortunately many political parties do not displays supporters in their social media main page. The political parties seemed forget that the supporters are one of the most important aspects that they should display in their social media main pages.

### 3.7 Social Media as Political Virtual Society

The potential of social media lies mainly in their support of civil society and the public sphere (Shirky, 2011). On the other hand, (Tumasjan, et al., 2010) results indicate that people are finding interesting political information on Twitter which they share with their network of followers. When they share any link of tweets to their network, then they automatically create a virtual society for their interest. Social media offer users new channels for political information (Kushin & Yamamoto, 2010).

Author visited every political parties' facebook account. Author found that every political party has different way to present their existness via facebook. Table 7 show the summary of political parties appereance in facebook.

Fig. 6. The most popular page based on SocialBakers

SocialBekrs reported that per April 2014, the most popular page in Indonesian Facebook was 1) Prabowo Subianto (fig. 1), followed by 2) Susilo Bambang Yudhoyono, and 3) Partai Gerakan Indonesia Raya.

Fig. 7. The fastest growing Facebook page in Indonesia

Among the most popular Facebook page, there are five fastest growing pages, 1) Prabowo Subianto, 2) Moeldoko, 3) PDI Perjuangan, 4) Komisi Pemberantasan Korupsi, and 5) Partai Gerakan Indonesia Raya (Gerindra). It means three of the fastest growing pages are related to political parties. The fastest person page is Prabowo Subianto, the head of PGerindra, as well as the presidential candidate of PGerindra. And the fastest political party's page is PDIP, the winner of 2014 election.

## 4 CONCLUSIONS

Based on the facts and discussions from previour sections above, author can conclude the following summarization related to the condition of social media related to general legislative elections results in Indonesia as follow:

1) Social media has positive impacts on political parties in Indonesian general election 2014. Social media is affecting political campaigns (Smith, 2011). National political parties in Indonesia already used some of social media for their political campaigns. Those social media are : 1) Facebook, 2) Twitter, and 3) YouTube. Facebook like is the symbols of popularity in Indonesian political athmosphere folowed by Twitter and YouTube.
2) Social media is effective tool for political cyber campaigns. The power of social media has triggers transparancy and support e-democracy around the world. Citizens have ability to choose freely the best legislative candidate to represent them in the parliament. In Indonesia, PGerindra won the third place in the real count, PGerindra also become political parties rallied highest voice revenue compared to the previous election. PGerindra also won in : 1) Facebook's likes, 2) Twitters' followers, and 3) YouTube' subscribers.
3) Social media in Indonesian political parties are dominated by young adults (18-24) followed by adults (25-34) users. This research confirm the engagements of young adults (Baumgartner & Morris, 2010).
4) Social media will create a more successful campaign as well as help create a stronger democracy (Vonderschmitt, 2012). There is no wall for every body to search the best candidate through online social media.
5) Social media are the current and future media for political campaigns and reach the voters and supporters instantly (Abdillah, 2014). Political parties are encourage to provide more professional social media pages over the internet and reach more online voters. Political parties' logos is the most common icon found in political parties' social media in Indonesian legislative general elections 2014 (Abdillah, 2014) followed by Quote, Slogan, and TagLine.
6) Information on Twitter can be aggregated in a meaningful way (Tumasjan, et al., 2010). Total followers have linear correlation with the voters in real election. In the case of Indonesia, PGerindra and PDIP success to get many voters. Contra conditions are faced by PGolkar and PKS. PGolkar used less effort in social media campaigns. PGolkar still works with traditional media, television, and they success to keep their loyal audiens. Another contra condition is faced by PKS, even they already force all of popular social media for political campaigns, the acquisition constituents that they obtained is lower than it should be.
7) In Obama's campaign, YouTube videos were successful in mobilizing younger voters like never before (McKinney & Rill, 2009). But, in Indonesian legislative campaigns, YouTube engaged less subcribers in every Indonesian political parties.
8) For next study, author interested to explore the power of social media in poilitics combined with blogs as digital democracy (Gil de Zúñiga, Veenstra, Vraga, & Shah, 2010), social media's events that makes Facebook the most used social network is being able to create an event (Curran, Morrison, & Mc Cauley, 2012), and related to presidential campaigns.

**Leon A. Abdillah** is a senior lecturer at Bina Darma University. He received his S.Kom. (BSc) and M.M. (MBA) degrees in computer science (information systems) and management (information systems) from the STMIK Bina Darma and Bina Darma University in 2001 and 2006, respectively. He ever continued his PhD study in computer science (information retrieval) in The University of Adelaide, South Australia (2010-2012). He is the author of dozens of scientific manuscripts, including: academic books, conference papers, journal articles, posters, and reports related to computer science, information technology, or information systems. He also reviewer/editor for dozens of national and international scientific journals, and dozens of national and international conferences. His current research interests include social media, information systems and databases, knowledge management systems, eLearning, and information retrieval. He is a member of APTIKOM, Balitbangda, IACSIT, IAENG, ICT Volunteer, IET, IKAPI, ISAET, SCIEI, SDIWC, and UACEE.